\documentclass[11pt,twoside]{article}

\usepackage{asp2006}
\usepackage{epsf}
\usepackage{lscape}

\begin{document}
\title{Mapping Mass in the Local Universe}   
\author{Karen L. Masters}   
\affil{Harvard-Smithsonian Center for Astrophysics, 60 Garden Street, Cambridge, MA 02138.}    

\begin{abstract}
We only see a small fraction of the matter in the universe, but the rest gives itself away by the impact of its gravity. Peculiar velocities have the potential to be a powerful tool to trace this matter however previous peculiar velocity surveys have struggled to meet their potential because of the large errors on individual measurements, poor statistics and uneven sky coverage. The 2MASS Tully-Fisher (2MTF) survey will make use of existing high quality rotations widths, new HI widths and 2MASS (2 Micron All-Sky Survey) photometry to measure Tully-Fisher  distances/peculiar velocities for all bright inclined spirals in the 2MASS redshift survey (2MRS). This survey based on the 2MASS galaxy catalog will provide a qualitatively better sample. It will provide significant improvements in sky coverage especially near the plane of our Galaxy which crosses the poorly understood "great attractor" region. I will give a progress report on the 2MTF survey including a look at over 300 hours of HI observations from the Green Bank Telescope (GBT) and a report on ongoing southern hemisphere observations with the Parkes Radio Telescope. The new spiral I-band field (SFI++) sample is currently the best available peculiar velocity survey for use in the local universe. I will also report on some preliminary results from this sample.
\end{abstract}

\section{Introduction}   

Despite the fact that cosmology has entered a new era with the concordance model
 of $\Lambda$CDM, several major questions remain, for example, the amount and distribution of dark matter and its relationship to the distribution of light in the Universe is still relatively uncertain. Such questions drive our understanding of galaxy and large scale structure formation and evolution. A number of galaxy redshift and distance surveys have already been used to address such questions,
however significant improvements could be made to results in the local universe by the use of a uniform, all-sky map combining both galaxy redshifts and distances.
   
In the concordance model visible matter makes up less than 4\% of the energy density of the universe or $\sim 10$\% of the gravitating matter, lighting up the ``cosmic web" like light fairy lights on a tree. The link between the growth of structure and the formation of galaxies (producing the visible light) is still not well understood and depends on the details of complicated baryonic physics. Cosmological simulations of the growth of structure model only the dark matter -- so the simplest way to check these models is to map the distribution of mass independent of the light. To do this we can trace it using the influence of its gravity. 

      In an expanding universe filled with gravitating matter there is an eternal battle between the expansion, which pulls everything apart and gravity pulling masses back together. Small density perturbations in the early universe, creating tiny regions where gravity begins to win out over the expansion, are able to grow and eventually collapse to form the structures we see in the universe today. The motions of galaxies towards these overdense regions is known as their {\it peculiar velocity}. In the linear regime (where both the deviations from a smooth universe and the velocities are small) the peculiar velocities of galaxies, ${\boldmath v}_{\rm pec}$, are directly proportional to the underlying gravitational  acceleration, ${\boldmath g}$. Furthermore the proportionality constant depends only on cosmological parameters, specifically Hubble's constant, $H_0$, the density of matter, $\Omega_M$, and the rate of growth of structure which can be aproximated as $f \sim \Omega_M^{0.6}$. Peculiar velocities can therefore be used to reconstruct the density field of the universe (if the cosmological parameters are known). Alternatively they can be used to measure cosmological parameters if the density field can be estimated ({\it e.g.} from galaxy redshift surveys). 
      
        Observationally only the radial component of the peculiar velocites can be measured from the difference between the observed recessional velocity of the galaxy and that expected for it using Hubble's Law and its known distance $v_{\rm pec} = v_{\rm obs} - H_0 d$. A typical peculiar velocity is on the order of a few hundred km~s$^{-1}$ -- the best known peculiar velocity is that of the local group which has a magnitude of $\sim 600$ km/s. Peculiar velocities for other galaxies can only be measured in the local universe since the error on an observed peculiar velocity is entirely dominated by the error on the measured distance. At $v_{\rm obs} = 1000$ km~s$^{-1}$ a 10\% error on the distance corresponds to a 100 km~s$^{-1}$ error on the measured peculiar velocity, but at $v_{\rm obs} = 10000$ km~s$^{-1}$ this error has inflated to 1000 km~s$^{-1}$, larger than the expected magnitude of most peculiar velocities. 
   
   Studies of peculiar velocities are further complicated since what we want to find is the vector field of peculiar velocities ${\boldmath v}$, while what we are able to measure is the radial component of that field, and with a large error (150-1000 km~s$^{-1}$). In theory it is possible to reconstruct the full 3D velocity field from measurements of the radial component since the velocity is proportional to the gravitational acceleration with is the gradient of a potential field. Reconstruction methods (eg. POTENT; Dekel et al. 1990) have been based on this principle, but in practice the large errors and uneven sky coverage of the velocity field measurements make the reconstruction very difficult.

 An understanding of peculiar velocities is not only interesting in itself but is also necessary to estimate distances to galaxies in the local universe where the peculiar velocity makes up a significant fraction of the observed recessional velocity. There are several thousand galaxies in the volume of space in which peculiar velocities are significant, only a few hundred of which have measured redshift-independent distances. So to learn about the physical properties of nearby galaxies, or to construct local luminosity/mass functions knowledge of the local peculiar velocity field is essential. 
    
\section{2MTF - The 2MASS Tully-Fisher Survey}   

 Previous peculiar velocity surveys have struggled to meet their potential because of large errors on individual measurements, poor statistics and uneven sky coverage. The 2MTF survey, being based on a source list from the 2 Micron All-Sky Survey (2MASS) will provide better statistics and more even sky coverage -- inparticular greatly reducing the impact of the zone of avoidance (the part of the sky inaccessible due to dust in our own Galaxy). This survey will make use of existing high quality rotations widths, new HI widths and 2MASS photometry to measure Tully-Fisher (TF) distances for all bright inclined spirals in the 2MASS Redshift Survey (2MRS; Huchra et al. 2004, 2005). 
 
 2MRS has now obtained redshifts of 95\% of galaxies in the 2MASS Extended Source Catalog (XSC; Cutri et al. 2003, Jarrett 2004) to K$_{\rm s}$=11.75 mag and $|b|>10^\circ$, ($\sim$ 40000 galaxies) and ultimately aims to be complete to K$_{\rm s}\sim$ 12.2 mag, $|b|>5^\circ$.  To a magnitude limit of K$_{\rm s}$=11.25 mag, $cz<10 000$ km/s, and axial ratio $b/a < 0.5$ 2MRS has $\sim$6000 galaxies. Over the whole sky, roughly 40\% of these objects have published rotation widths suitable for use in TF, but the sky distribution of available widths is very uneven.
  
The largest sources of rotation widths are the HI archive of Springob et al. (2005), and the optical rotation curve (ORC) database maintained by Haynes \& Giovanelli. While HYPERLEDA ({\it e.g.} Paturel et al. 2003) catalogs a large number of HI widths, many do not have sufficient $S/N$ or velocity resolution for TF (although new widths are coming from the KLUN+ project at Nancay, e.g. Theureau et al. 2007).
On a timescale of 3--5 years, the ALFALFA survey (Giovanelli et al. 2005) will provide high resolution widths for all HI rich galaxies in the high $|b|$ Arecibo sky to $S_p\sim10$ mJy.  The HIPASS survey in the south (Barnes et al. 2001)  does not have sufficient velocity resolution for TF and has a rather shallow HI flux limit.

 New observations are required to take advantage of the superior sky coverage provided by selection from 2MRS, particularly in the southern hemisphere and north of the declination range accessible by Arecibo.  In this effort we have been successful in the past year obtaining telescope time at both the Green Bank Telescope (GBT) and the Parkes Radio Telescope. The current status of the 2MTF project is shown in Figure \ref{masters:2mtf}. Details of our observing over the past year are given below.

\begin{figure}
\plotfiddle{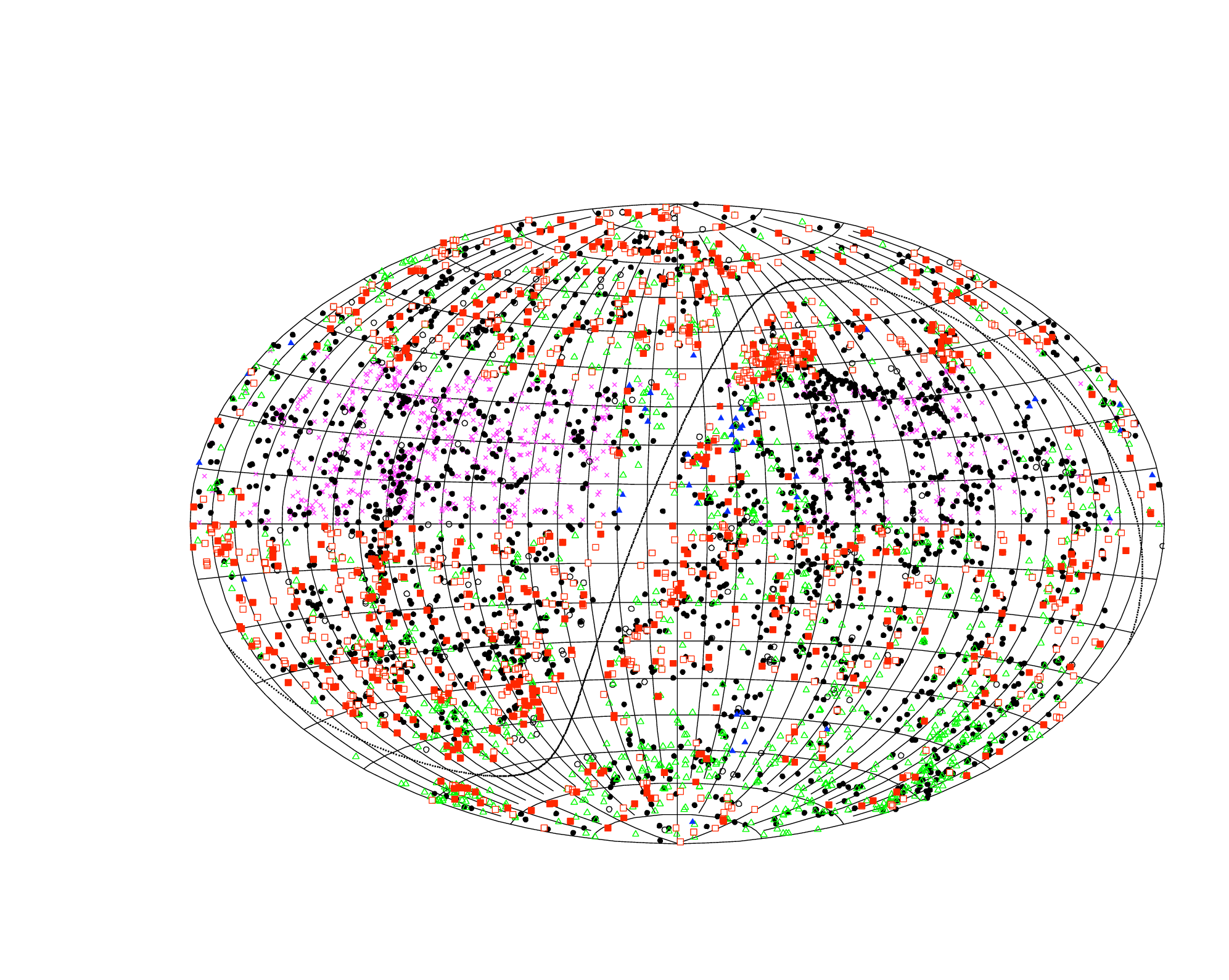}{2.9in}{0.}{75}{75}{-230}{-47}
\caption{Sky distribtion of galaxies in the 2MTF sample. Filled/open circles - galaxies with published widths, filled squares - galaxies with widths from our GBT and Parkes observations. Open squares - poor/non-detections from our observations. Open triangles - poor candidate for observing (early type spirals). Filled triangles - galaxy not yet observed. (Crosses - no published width in ALFALFA sky area).\label{masters:2mtf}}
\end{figure}

\subsection{GBT Observing}
\begin{figure}
\plotone{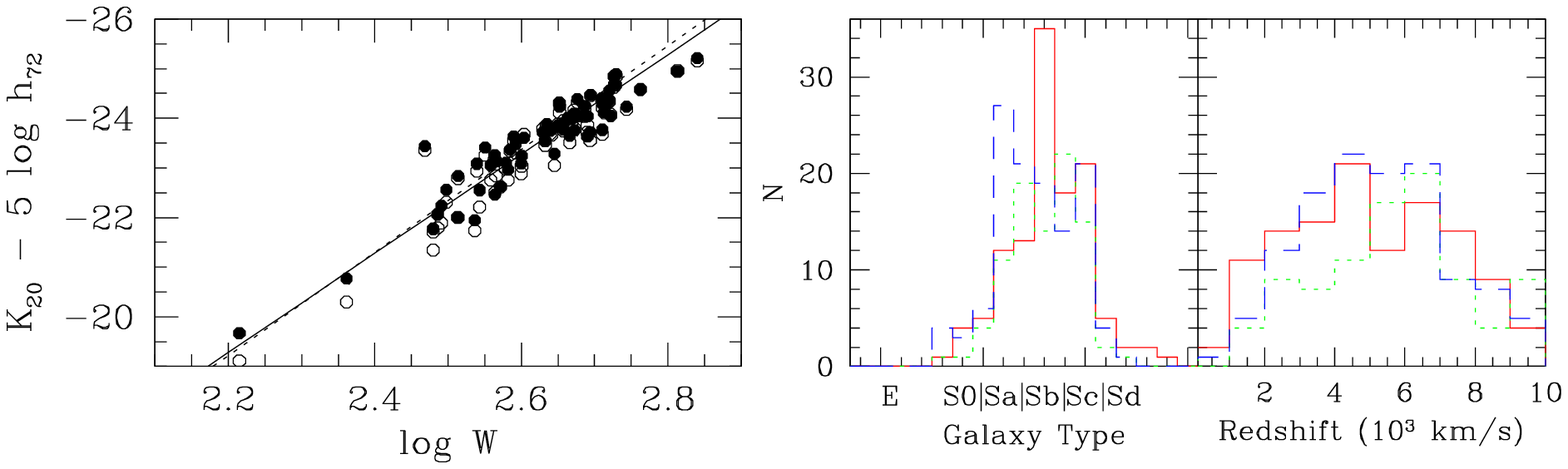}
\caption{On the left is K-band absolute magnitude vs. raw rotation width (corrected for inclination) for the first 70 galaxies with high quality HI profiles from our GBT observations. Solid points show magnitudes using distances from the multiattractor model of Masters (2005), while open points use Hubble's Law. The solid line shows a TF relation of $M_{\rm K} - 5 \log h = -22.28 - 10 (\log W - 2.5)$, for which the zeropoint is a fit to multiattractor model corrected magnitudes. The dotted line shows the Cepheid calibration of the K-band TF from Macri (2001). On the right are shown the distribution of galaxy type and redshift for all galaxies observed in the first semester of our GBT observing. The solid line shows good detections, the dotted line shows galaxies detected, but with insufficient S/N for TF. The dashed line shows non-detections to a limit of $S_p \sim 5$mJy. \label{masters:GBT}}
\end{figure}

\begin{figure}
\plotfiddle{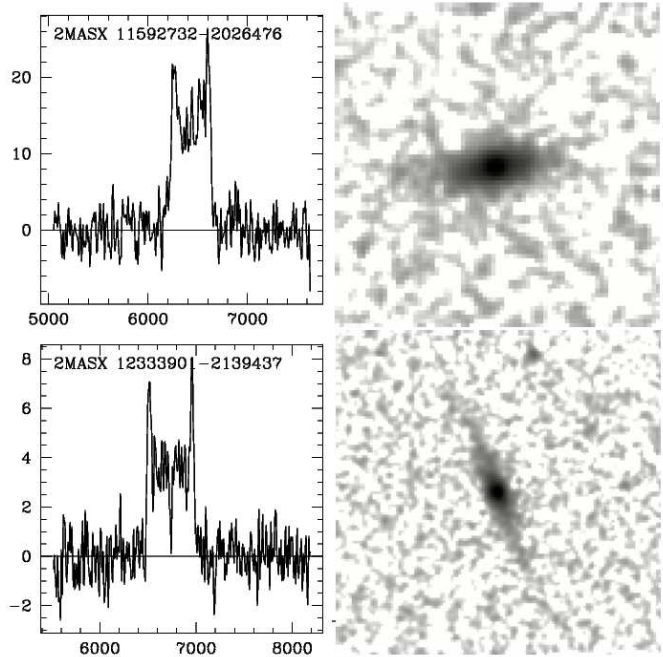}{3.3in}{0.}{50}{50}{-160}{-30}
\caption{Example HI line profiles from our GBT observations plotted next to the 2MASS K-band Image of the respective galaxy. \label{masters:gbtprofiles}}
\end{figure}

 In 2006 we were granted 381 hours of Green Bank Telescope (GBT) time in which we  observed 996 galaxies in the GBT accessible sky (north of Dec=$-40^\circ$) to a rough peak flux limit of $S_p \sim 10$ mJy.  Since the source list came from 2MRS redshifts were already known for all targets. Observations were done in postion-swithed mode normally with integration times varying from 5-25 mins ON/OFF per galaxy. We detected HI from 606 galaxies (61\%), 410 (68\%) of which have sufficient S/N for Tully-Fisher. Figure \ref{masters:GBT} shows a preliminary TF relation with the first  few reduced widths and some observing statistics from the first semester of observing. Figure \ref{masters:gbtprofiles} shows two example profiles next to the K-band 2MASS image of the corresponding galaxy. Data reduction is in progress using the GBTIDL software with a catalogue release anticipated later this year. 

\subsection{Parkes Observing}
This year we were also granted 256 hours of observing time at the Parkes Radio Telescope. In this time we observed 149 galaxies in the extreme South (DEC $< -40^\circ$) to a peak flux limit of $S_p \sim 20$ mJy. To this limit HIPASS spectra (Barnes et al. 2001) can be used as a quick first look. Galaxies at these fluxes are not commonly catalogued HIPASS sources but with {\it apriori} knowledge of the galaxy redshift the signal can sometimes be picked out of the HIPASS spectra. We detected HI in 107 (72\%) of the galaxies, 90 (84\%) of which have sufficient S/N for TF. We recently received a favourable review and will likely be schedule for more time to continue this work in the 07OCTS observing semester at Parkes.

\section{Early Results from SFI++ Peculiar Velocities}

 SFI++ combines I-band imaging with HI or ORC widths to measure TF distances/peculiar velocities for close to 5000 spiral galaxies distributed over the whole sky (Springob et al. 2007). SFI++ galaxies are selected optically so the sky distribution is not as good as it will be for 2MTF, however it is the largest peculiar velocity catalogue currently available from a single source and as such the SFI++ sample is the best currently available for studies of the local velocity field. The full SFI++ data is available for download from \\ \texttt{http://arecibo.tc.cornell.edu/hiarchive/sfiplusplus.php}. 
 
\begin{figure}
\plotfiddle{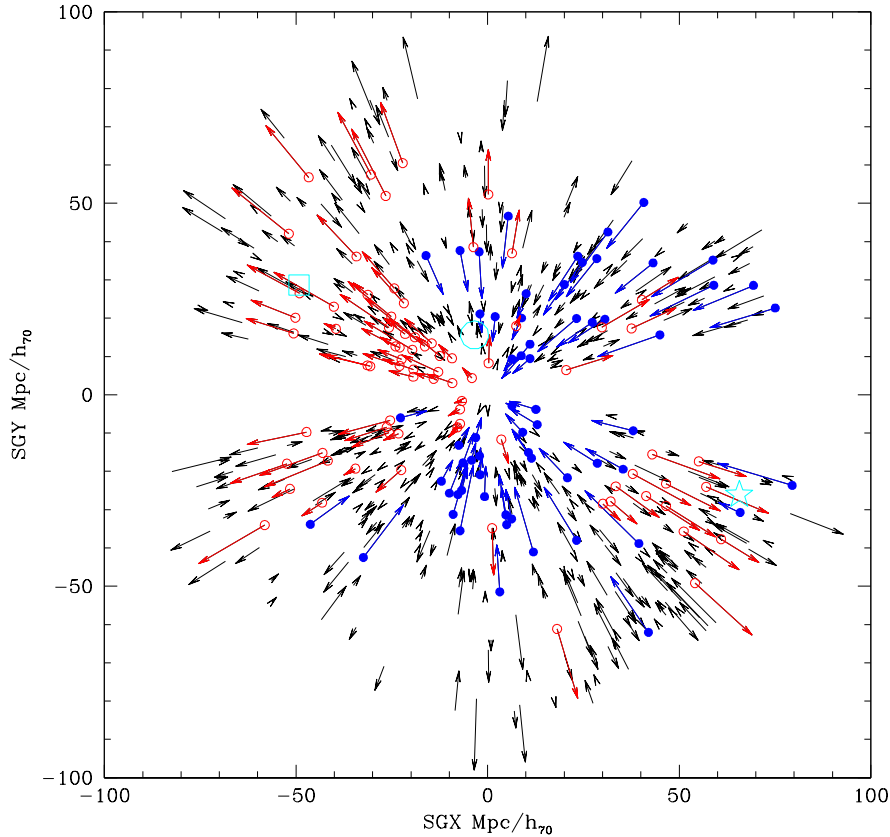}{4.3in}{0.}{60}{60}{-185}{-100}
\caption{SFI++ peculiar velocities are shown for objects within 15 Mpc of the supergalactic plane (SGP); velocities are shown projected onto the SGP. If the velocity has a significant ($> 1\sigma$) positive or negative value it is shown with a open or filled circle respectively. The large open circle indicates the location of the Virgo Cluster, the square shows the center of the Hydra-Centaurus supercluster and the star indicates Pisces-Perseus (positions as given in Einasto et al. 1997). \label{masters:sfi++}}
\end{figure}

  Several interesting results have already been derived from the SFI++ sample, in particular during the construction of the I-band TF template relation (Masters et al. 2006). A simple multiattractor model to describe the basic features of the local velocity field (and for the estimation of local distances from redshifts) has also been fit to the data (Masters 2005; Masters in prep.). The local velocity field from SFI++ (in the plane of the Local Supercluster) is shown in Figure \ref{masters:sfi++}
  
 \subsection{The Scale of Density Fluctuations}
 The motions of clusters probe the peculiar velocity field on a scale which is well within the linear regime where theory and observation can be linked relatively easily. Large numbers of galaxies in each cluster also allow the peculiar velocities of clusters to be calculated to relatively high accuracy. Cluster peculiar velocities therefore have great potential to provide constraints on cosmological parameters.  As explained in Watkins (1997), the 1D peculiar velocity dispersion of clusters depends relatively simply on the scale of density fluctuations and the total mass density of the universe -  a relationship coming from the dependence on the initial power spectrum of fluctuations. In a flat universe ($\Omega_M + \Omega_\Lambda = 1$) the relation is given by
\begin{equation}
\sigma_v = \frac{100 {\rm km s}^{-1}}{\sqrt{3}} \Omega^{0.6} \sigma_8 \sqrt{f(\Omega h)}
\end{equation}
where $\sigma_8$ is as usual the $rms$ fluctuation on scales of 8 Mpc/$h$, $\Omega$ is the mass density of the universe, and $h = H_0/100$ km/s Mpc$^{-1}$. In Watkins (1997) the function $f(\Omega h)$ is well fit by $f = 12.5 (\Omega h)^{-1.08} + 49.4$, and the approximation $\sqrt{f} \sim 10$ over the range $\Omega h = $ 0.2--0.5 is used. The third year WMAP data release quotes $\Omega h^2 = 0.134\pm0.006$, $h=0.73\pm0.03$ and $\sigma_8 = 0.84\pm0.06$ (Spergel et al. 2007), implying a value for the velocity dispersion of $\sigma = 239\pm23$ km/s. 

 In Masters et al. (2006) peculiar velocities are derived for 31 clusters which are used in the construction of the SFI++ template TF relation. The raw velocity dispersion from these clusters is  $\sigma = 365\pm34$ km/s. This measurement is biased larger than the intrinsic cluster peculiar velocity dispersion because of broadening due to the error on the individual cluster peculiar velocity measurements. Correcting for error broadening as estimated from a Monte Carlo simulation gives $\sigma = 298 \pm 34$km/s.
 
This measurement of the cosmic cluster peculiar velocity dispersion is consistent with the WMAP estimate, providing independent verification of the best fit cosmological model. We can use the approximation $\sqrt{f} \sim 10$ in Equation (1) to calculate a value $\Omega^{0.6} \sigma_8 = 0.52\pm0.06$ from this (where the error here only represents the measurement error on the cluster peculiar velocity dispersion and does not account for any bias introduced by the assumption). Alternatively we can use WMAP information on $\Omega h^2=0.134\pm0.006$ (coming from the relative amplitude of the power spectrum peaks) along with the measurement of $h=0.74\pm0.02$ (see below) from a combination of TF and Cepheid data to calculate $\sqrt{f}=11.34\pm0.59$, implying $\Omega^{0.6} \sigma_8 = 0.46\pm0.06$.
 
 \subsection{Hubble's Constant}
 In Masters et al. (2006) a classic ``distance ladder" approach is used to measure a value for the Hubble constant from a combination of Cepheid and TF distances. Cepheid distances to SFI++ galaxies are used to calibrate the TF zeropoint. Distances to the set of clusters used to define the slope of the TF relation are then used to calculate the Hubble constant. The bulk of the Cepheid distances for SFI++ galaxies come from the HST Key Project (Freedman et al. 2001). For more details see Masters et al. (2006). We measure $H_0 = 74\pm2$ (random) km/s Mpc$^{-1}$. We estimate that a further systematic error of $\pm 6$ km/s Mpc$^{-1}$ is introduced by the uncertainty in the calibration of the Cepheid distances. 

 This measurement of $H_0$ is identical to the determination of $H_0 = 74\pm2$ km/s Mpc$^{-1}$ (S\'{a}nchez et al. 2006) from a combination of WMAP and 2dFRGS data. Other measurements of $H_0$ using Cepheid calibrations of the TF relation include the HST Key Project value of $H_0=71\pm4\pm7$ km/s Mpc$^{-1}$ ~(Sakai et al. 2000), and the SFI measurement of $H_0 = 69 \pm 2 \pm 6$ km/s Mpc$^{-1}$~(Giovanelli et al. 1997; note that neither of these determinations accounted for the metallicity dependence of the Cepheid zeropoint as was done in Masters et al. 2006). The dominant source of error in this determination of the value of $H_0$ comes not from the TF relation, but from the uncertainty in the zeropoint calibration of the Cepheid relation, and ultimately from the uncertainty of our knowledge of the distance to the LMC. If this error were not present TF could measure $H_0$ to a comparable accuracy to the best ``precision cosmology'' available, and better than is possible from just WMAP alone which quotes $H_0 = 73 \pm 3$ km/s Mpc$^{-1}$ ~and also requires an assumption that the universe is flat (Spergel et al. 2007).

\subsection{A Multiattractor Model for the Velocity Field}
Numerical simulations of the growth of structure in a $\Lambda$CDM universe have shown that motion along filaments towards the elliptical dark matter halos of clusters form the dominant pattern of infall in the universe. However a simple multiattractor model (describing infall onto one or more spherical attractors) can still provide insight into the dominant sources of attraction in the local volume. It also provides a useful parametric correction to distances derived from redshifts nearby. The best fit model for the SFI++ peculiar velociites includes infall onto Virgo and the Hydra-Centaurus supercluster, a quadrupolar component describing smaller than average expansion out of the supergalactic plane and a residual dipole pointing towards the Hydra-Centaurus/Shapley superclusters. The motion of the Local Group can be completely explained by this model, leaving no evidence for a ``Local Anomaly'' (excess LG motion towards -SGZ) which has previously been suggested (Masters 2005)

 In Masters (2005) it is argued that this model provides evidence for infall onto both the Hydra-Centaurus and Shapley superclusters. Infall onto Shapley, which is significantly outside the sample volume is best fit as a bulk flow and accounts for at least part of the residual dipole. Previous arguments about the nature of the ``great attractor'' can be solved by this physically obvious result. In particular, the mass of the nearby ``great attractor'' associated with Hydra-Centaurus is most likely inflated in models which do not also account for infall onto the Shapley supercluster and is therefore not as significant as suggested in previous ``great attractor'' models. 

The parametric velocity field presented in Masters (2005) is fit to a sample of velocity field tracers almost an order of magnitude larger than any that has been used before. The SFI++ sample should also be exploited for more complex non-parametric models of the velocity field, where the larger numbers along with high quality peculiar velocities will provide higher resolution views of the smoothed density field than was possible with previous samples. Meanwhile, this parametric modeling provides the best currently available correction for redshift distances in the local universe.

\section{Conclusions}
 The peculiar velocities of galaxies have the potential to provide a powerful tool to trace the distribution of mass in the universe independent of the complicating details of galaxy and star formation. However previous peculiar velocity studies have struggled to meet their potential because of uneven sky coverage and the large errors associated with individual measurements. 
 
 We present plans for a new peculiar velocity survey - the 2MASS Tully-Fisher (2MTF) survey. This survey, based on selection from the 2MASS galaxy catalogue will have significantly more even sky coverage than has been available before - in particular near the plane of our Galaxy. It will use 2MASS photometry and both published and new rotation widths to measure TF distances/peculiar velocities for all inclined spirals in 2MRS. This survey will provide a {\it qualitative} improvement on what has been available up to now. We provide a progress report on 381 hours of GBT observations and 256 hours at the Parkes Radio Telescope as part of this effort.
 
  We present early results from the SFI++ peculiar velocity sample -- the best peculiar velocity sample currently available to study the local universe (Springob et al. 2007). This survey has been used to measure the cluster peculiar velocity dispersion (a quantity which depends simply on the scale of density fluctuations and the matter density in the universe) to be $298\pm34$ km/s, which implies  $\Omega^{0.6} \sigma_8 = 0.52\pm0.06$ (Masters et al. 2006). SFI++ data have also been used (in combination with Cepheid distances to several SFI++ galaxies) to measure $H_0 = 74\pm2$ (random) $\pm6$ (systematic) km/s Mpc$^{-1}$ -- a precision comparable with the best available measures for $H_0$ (Masters et al. 2006). Finally a simple multiattractor model has been fit to the SFI++ data in the local universe. This model gives insight into the dominant sources of attraction nearby and also provides a useful parametric correction to distances derived from redshifts nearby (Masters 2005; Masters in prep.).
  
   The TF relation still has a place in the era of ``precision cosmology'' to provide independent checks of the best fit cosmological model. It has the potential to measure some parameters (notably the cluster peculiar velocity dispersion and $H_0$) to an accuracy comparable to the best available. Peculiar velocities also provide a technique which should not be discarded (for all it's difficulties) to study the {\bf mass} distribution locally and to learn about the basic cosmology of our universe.
 
\acknowledgements KLM is a Harvard Postdoctoral Research Fellow supported by NSF grant AST-0406906. GBT is part of the National Radio Astronomy Observatory which is a facility of the National Science Foundation operated under cooperative agreement by Associated Universities, Inc. The Parkes telescope is part of the Australia Telescope which is funded by the Commonwealth of Australia for operation as a National Facility managed by CSIRO. 2MASS Atlas Images obtained as part of the Two Micron All Sky Survey (2MASS), a joint project of the University of Massachusetts and the Infrared Processing and Analysis Center/California Institute of Technology, funded by the National Aeronautics and Space Administration and the National Science Foundation.

I wish to thank all collaborators on the 2MTF project who have helped with GBT and Parkes observing, especially Aidan Crook and Lucas Macri. I also wish to thank numerous collaborators and members (past and present) of the Cornell Extragalactic group without whose work the SFI++ sample would not have been possible.

\end{document}